\documentclass[twocolumn,showpacs,superscriptaddress,preprintnumbers,amsmath,amssymb,prc]{revtex4}

\usepackage{graphicx}
\usepackage{dcolumn}
\usepackage{bm}
\usepackage{float}
\usepackage{color}
\usepackage{CJK}
\usepackage{ulem}
\usepackage[colorlinks,linkcolor=blue,urlcolor=blue,citecolor=blue]{hyperref}

\begin{document}

\begin{CJK*}{UTF8}{gbsn}

\title{
Light nuclei production in Au + Au collisions at $\sqrt{s_{NN}}$ = 7.7 - 80 GeV from UrQMD model}
\author{Xian-Gai Deng(邓先概)}
\affiliation{Key Laboratory of Nuclear Physics and Ion-beam Application (MOE), Institute of Modern Physics, Fudan University, Shanghai 200433, China}
\author{Yu-Gang Ma(马余刚)\footnote{Corresponding author: mayugang@fudan.edu.cn}}
\affiliation{Key Laboratory of Nuclear Physics and Ion-beam Application (MOE), Institute of Modern Physics, Fudan University, Shanghai 200433, China}
\affiliation{Shanghai Institute of Applied Physics, Chinese Academy of Sciences, Shanghai 201800, China}

\date{\today}

\begin{abstract}

Light nuclei production in relativistic $^{197}$Au + $^{197}$Au collisions from 7.7 to 80 GeV is investigated within the Ultra-relativistic-Quantum-Molecular-Dynamics model (UrQMD) with a naive coalescence approach. The results of the production of light nuclei at midrapidity can essentially match up the experimental data and a slight  enhancement of combined ratio of ${N_{p}N_{t}}/{N_{d}^{2}}$ where $N_p, N_d$ and $N_t$ represent respectively the yields of proton, deuteron and triton, which is sensitive to the neutron density fluctuations, occurs around 20 GeV.  However,  this enhanced ${N_{p}N_{t}}/{N_{d}^{2}}$ ratio should not be over-understood considering that the present UrQMD model is a  cascade version without equation of state (EoS), i.e. there is an absence of critical end point mechanism.  Furthermore,  within different rapidity regions, the kinetic temperatures of different light nuclei are extracted by the Blast-wave model analysis and ratios among different light nuclei  are  also discussed. 

\end{abstract}


\maketitle

\section{Introduction}
\label{introduction}

One of goals in relativistic heavy-ion collisions is to explore the phase diagram of Quantum Chromodynamics Dynamics (QCD). The conjectured QCD phase diagram which can be expressed as a plot of temperature vs baryon chemical potential (T, $\mu_{B}$) has been several decades from the first drawing \cite{CN75,GB02,KF08,KF11,LXF17}. One of features in the QCD phase diagram is so-called the critical end point (CEP) \cite{RA07} of the first order phase transition from hadronic phase to quark-gluon phase in this diagram which was first proposed in 1989  \cite{MA89,AB89}. This is a current challenge both from experimental and theoretical sides. There are many techniques so far to search for  the location of this critical end point  in phase diagram, such as the lattice calculations \cite{ZF02,PF03}, the ratio of viscosity to entropy density ($\eta/s$) \cite{DT03,SG06}, cumulants (skewness and kurtosis) \cite{MA11,HC18,JM19}, conserved charge and baryon density fluctuations  \cite{LXF17,CS07,MA09-0} as well as higher order moment \cite{MA09-1,JS12,Ko16,LF17} etc, however, no consensus was reached yet.  Recently, as proposed by Ref.~\cite{KJ17,KJ18} based on coalescence model as well as by preliminary results from the STAR collaboration with the Beam Energy Scan (BES) program, one found that there exists a non-monotonic relation of the  ratio $N_{t}N_{p}/N_{d}^{2}$, which could be related to the neutron density fluctuation, as a function of center-of-mass energy $\sqrt{s_{\rm NN}} $ \cite{ZDW20,LH20} and it triggers  many interesting works on exploration of the ratios of light nuclei \cite{ES20,ES20b,LH20,KJ20}. 
In Ref.~\cite{KJ20b}, it is found that the first-order chiral phase transition can enhance the ratio of $N_{t}N_{p}/N_{d}^{2}$. 
These results suggest that realistic equation of state or CEP mechanism should be needed. In Ref.~\cite{JW20}, one calculated baryon probability density by UrQMD and found no  baryon density fluctuation as claimed by experimental indication. In this context,  lots of efforts are still needed on addressing non-monotonic issue. 

On the other hand,  only midrapidity region was focused in most experimental measurements as well as theoretical calculations so far, and less efforts are paid on productions of light-nuclei and their ratios in large rapidity regions. For central collisions at  a given energy, one can separate rapidity into various regions. If these various rapidity regions correspond to the various initial condition as $(T_{0}, \mu_{B0})$, then these initial conditions $(T_{0}, \mu_{B0})$ would have their own trajectory during cooling process. Thus we concern that the CEP could occur in an energy region with different rapidity windows. Therefore it is of interests to check the rapidity dependence of light-nuclei. Based upon the above arguments,  we have two main motivation in this work. One is to investigate density  fluctuations by the yield ratios of light nuclei in the midrapidity region, and another is to extract light-nuclei production and  ratios in higher rapidity regions and see what difference from the midrapidity ones.

\section{Model and methodology}
\label{TF1}
\subsection{UrQMD model and coalescence}

The UrQMD model is one of microscopic models and  extensively used in simulating the ultra-relativistic heavy ion collisions \cite{SA98,MB99,HP08,WangYJ}. 
The mean field potential is taken into account as  the collision c.m. energy $\sqrt{s_{\rm NN}}$ which is less than 3.3 GeV, however, the present simulations which are above 7 GeV are only with cascade part.  In UrQMD model, the degrees of freedom are hadrons and strings. The more details can be found in Refs.~\cite{SA98,HP08,WangYJ}.   In many other works, thermal and statistical approaches are used to describe the  production of light nuclei \cite{AA11,NS16}. Here by a coalescence mechanism with the final phase space information of baryons, we can obtain production yields of light nuclei.  For more details,  a light nucleus can be recognized by a so-called minimum spanning tree (MST) clusterization algorithm based on coordinate and momentum cuts,  which was also utilized to determine nuclear fragments in the Quantum Molecular Dynamics  simulations \cite{AJ91}. The yield of nucleus is given by the condition of  $\bigtriangleup r <$ 3.575 fm and  $\bigtriangleup p <$ 0.285 GeV$/c$ as adapted  in Ref.~\cite{SS19}, where $\bigtriangleup r$ ($\bigtriangleup p$) means the relative spatial distance  (momentum) between two particles in the two-particle rest frame at equal time. The stopping time for simulations is taken at 65 fm$/c$, which is long enough for the selected energy domain of Au + Au collisions.
 As discussed in Refs. \cite{SS19,LQ16-1,LQ16-2,LQ16-3}, there are some  effects on the light nuclei productions due to different selection of the values of  $\bigtriangleup r$ and $\bigtriangleup p$, we also test other combination of  $\bigtriangleup r$ and $\bigtriangleup p$, such as ($\bigtriangleup r <$ 3.00 fm, $\bigtriangleup p <$ 0.285 GeV$/c$) and ($\bigtriangleup r <$ 3.575 fm, $\bigtriangleup p <$ 0.35 GeV$/c$). It is found that the ($\bigtriangleup r <$  3.575 fm, $\bigtriangleup p <$ 0.285 GeV/c) is the better choice. Of course, more combinations of ($\bigtriangleup r$,$\bigtriangleup p$) could be tested for the best selection, but our selection of ($\bigtriangleup r$, $\bigtriangleup p$) was also based on the successful description to deuteron production  in previous work by the same model ~\cite{SS19}, then we can take  those values in following calculations.

For simplicity,  the spin and isospin factors are not yet considered in this work. The UrQMD-3.3p1 version is applied to simulate central  $^{197}$Au + $^{197}$Au collisions with impact parameters of b = 0 - 3 fm at $\sqrt{s_{\rm NN}}$ = 7.7 GeV to $\sqrt{s_{\rm NN}}$ = 80 GeV. 

\subsection{Ratios and density fluctuation}

Light nuclei are usually formed during cooling process of hot and dense medium and can then be used to extract important information of nucleon distributions at freeze-out. The coalescence of nucleons is related to the local nucleon density \cite{ST63,HH76,HS81,JA19}. In the coalescence model, the numbers of deuteron and triton can be given by \cite{KJ17,LH20}:
\begin{eqnarray}
\label{ND1}
 &N_{d}&= \frac{3}{2^{1/2}}(\frac{2\pi}{m_{0} T_{\rm eff} })^{3/2} \frac{N_{p}N_{n}}{V}, \\
\label{NT1}
 &N_{t} &= \frac{3^{3/2}}{4}(\frac{2\pi}{m_{0} T_{\rm eff}})^{3} \frac{N_{p}N_{n}^{2}}{V^{2}}, 
\end{eqnarray}
where $V$ is system volume and $T_{eff}$ is temperature of source at kinetic freeze-out. The $N_{p}$, $N_{n}$ and $m_{0}$ are proton number, neutron number and nucleon mass ($m_{p}=m_{n}$), respectively. One introduces density fluctuations of nucleons, 
\begin{eqnarray}
\label{RHN2}
 &\rho_{n}(\vec{r})&= \frac{1}{V} \int \rho_{n}( \vec{r}) d \vec{r}+ \delta \rho_{n}(\vec{r}) = \langle \rho_{n} \rangle + \delta \rho_{n}(\vec{r}), \\
\label{RHP2}
 &\rho_{p}(\vec{r})&= \frac{1}{V} \int \rho_{p}( \vec{r}) d \vec{r}+ \delta \rho_{p}(\vec{r}) = \langle \rho_{p} \rangle + \delta \rho_{p}(\vec{r}), 
\end{eqnarray}
where $\rho_{n}$ and $\rho_{p}$ are densities of neutron and proton, respectively; $\langle \cdots \rangle$ denotes average value. It can be rewritten 
Eq. ~(\ref{ND1}) and Eq. (\ref{NT1}) \cite{KJ17},
\begin{eqnarray}
\label{ND2}
&N_{d}&= \frac{3}{2^{1/2}}(\frac{2\pi}{m_{0} T_{\rm eff}})^{3/2} N_{p}\langle \rho_{n} \rangle(1+\alpha \triangle \rho_{n}), \\
\label{NT2}
 &N_{t} &= \frac{3^{3/2}}{4}(\frac{2\pi}{m_{0} T_{\rm eff}})^{3}  N_{p}\langle \rho_{n} \rangle[1+(1+2\alpha) \triangle \rho_{n}], 
\end{eqnarray}
 where $\triangle \rho_{n} = \langle (\delta \rho_{n})^{2} \rangle/\langle \rho_{n} \rangle^{2}$ is the relative density fluctuation of neutrons, and $\alpha$ is the correlation coefficient. Taking Eq. ~(\ref{ND2}) and Eq. ~(\ref{NT2}) into account, one can define a yield ratio,
\begin{eqnarray}
\label{RT1}
O_{1} = \frac{N_{p}N_{t}}{N_{d}^{2}}=g\frac{1+(1+2\alpha) \triangle \rho_{n}}{(1+\alpha \triangle \rho_{n})^{2}}
\end{eqnarray}
with $g$ = 4/9 $\times (3/4)^{1.5}\approx$ 0.29. One constructs the ratio $O_{1}$ as such a way, then one could remove the effects of volume $V$, temperature $T_{eff}$ and isospin asymmetry of the emission source.  When neutron and proton density fluctuations are completely correlated, $\alpha$ is equal to 1. Then one can  get,
\begin{eqnarray}
\label{RT2}
O_{1}\approx 0.29 (1+\triangle \rho_{n}).
\end{eqnarray}
From Eq.~($\ref{RT2}$), the light-nuclei ratio is relative to the neutron density fluctuation. And the density fluctuations would be amplified in the spinodal region of phase diagram \cite{JS12}.  Results in Ref.~\cite{KJ17} suggested that the yield ratio of light nuclei in relativistic heavy-ion collisions can be taken as a direct probe of the large density fluctuations which might be associated with the QCD critical phenomenon.

Moreover, other ratios of light-nuclei which involving $^{4}{\rm He}$ were proposed in Ref.~\cite{ES20,ES20b},
\begin{eqnarray}
\label{RT3}
&O_{2}&=\frac{N_{^{4}{\rm He}}N_{p}}{N_{^{3}\rm He}N_{d}}, \\
\label{RT4}
&O_{3}&=\frac{N_{^{4}{\rm He}}N_{t}N_{p}^{2}}{N_{^{3}{\rm He}}N_{d}^{3}} , \\
\label{RT5}
&O_{4}&=\frac{N_{^{4}{\rm He}}N_{p}^{2}}{N_{d}^{3}}. 
\end{eqnarray}
Since these above ratios have the same powers of fugacity in denominators and numerators, so they can cancel and eliminate the dependence of baryonic chemical potential. From the results in Ref. \cite{ES20,ES20b}, one suggests that these ratios are sensible indicators of critical behavior. Also, we found that only $O_1$ and $O_2$ is independent each other, but $O_3$ can be expressed by $O_1 \times O_2$, and   $O_4 = O_3/\frac{N_t}{N_{^3 {\rm He}}} $  In our simulations, some single ratios such as the ratios of neutron to proton ($N_n/N_p$), triton to $^{3}$He ($N_t/N_{^{3}{\rm He}}$), and $^{4}$He to $^{3}$He ($N_{^{4}{\rm He}}/N_{^{3}{\rm He}}$) are also considered.

\begin{figure}[htb]
\setlength{\abovecaptionskip}{0pt}
\setlength{\belowcaptionskip}{8pt}
\includegraphics[scale=1.0]{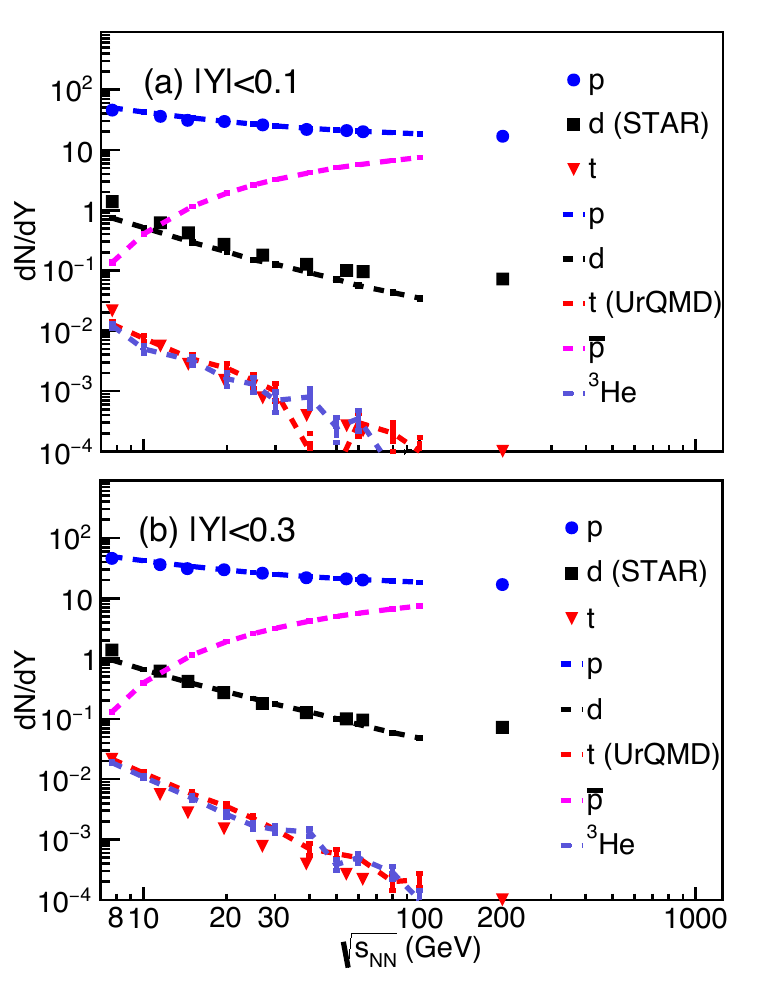}
\caption{(Color online) Yield of light nuclei as a function of c.m. energy $\sqrt{s_{\rm NN}}$ in different rapidity regions. Dash lines are from the central (b $<$ 3 fm) $^{197}$Au + $^{197}$Au collisions within the UrQMD model, and dots are the preliminary results from the STAR Collaboration with the midrapidity cuts $|$Y$|$$<$0.1, $|$Y$|$$<$0.3 and $|$Y$|$$<$0.5 for protons, deuterons and tritons, respectively \cite{BI09,LA17,JA19,LH20}.}
\label{fig:fig1}
\end{figure}

\begin{figure*}
\setlength{\abovecaptionskip}{0pt}
\setlength{\belowcaptionskip}{8pt}
\centerline{
\includegraphics[scale=0.58]{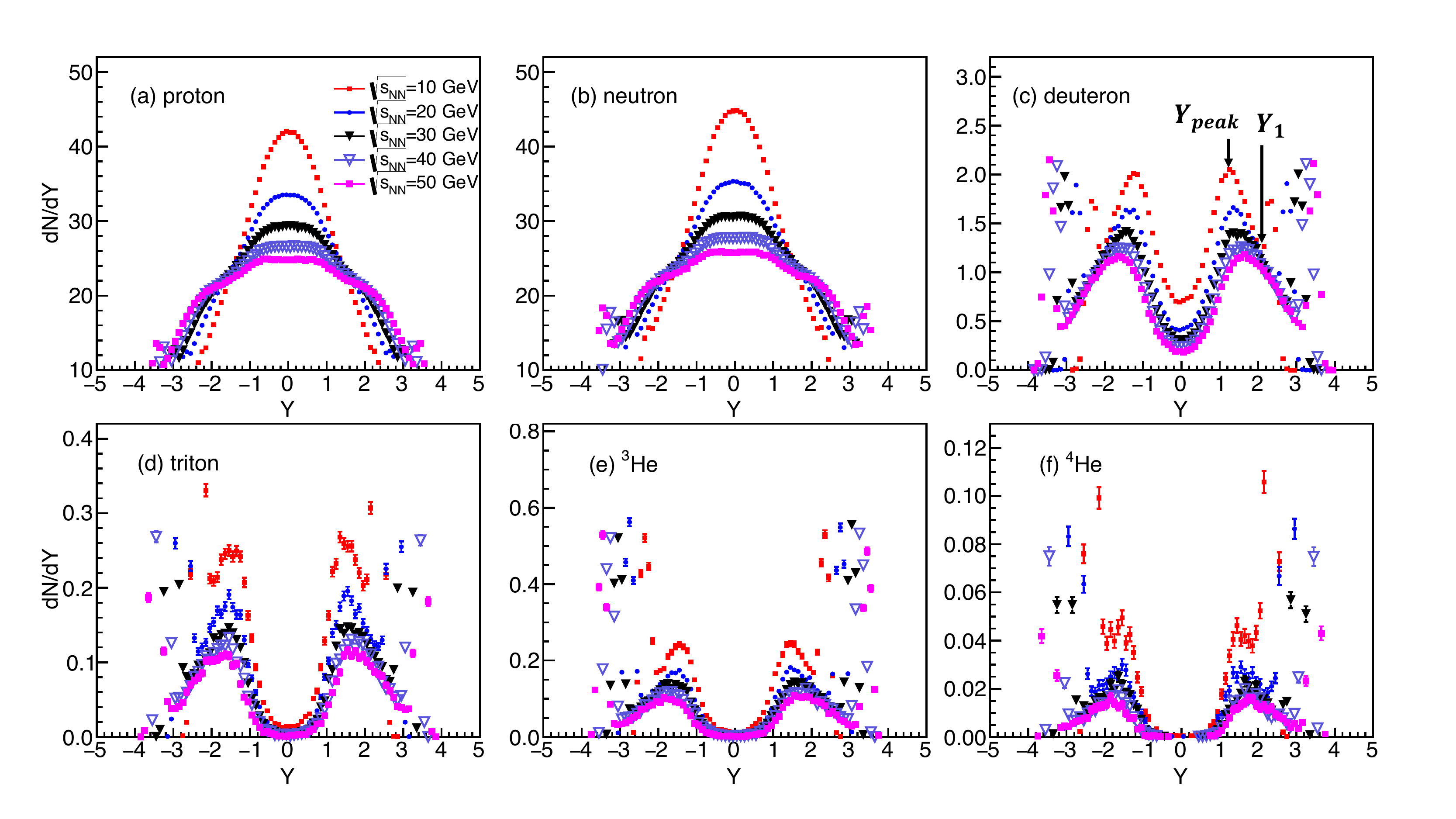}}
\caption{(Color online) Rapidity distributions of various light nuclei in central (b $<$ 3 fm) $^{197}$Au + $^{197}$Au collisions at different energies within the UrQMD model. In (c), $Y_{peak}$ and $Y_1$ indicates of the rapidity positions of first peak and valley in right-hand rapidity distribution at $\sqrt{s_{NN}}$ = 10 GeV. }
\label{fig:fig2}
\end{figure*}

\begin{figure}
\setlength{\abovecaptionskip}{0pt}
\setlength{\belowcaptionskip}{8pt}
\includegraphics[scale=0.56]{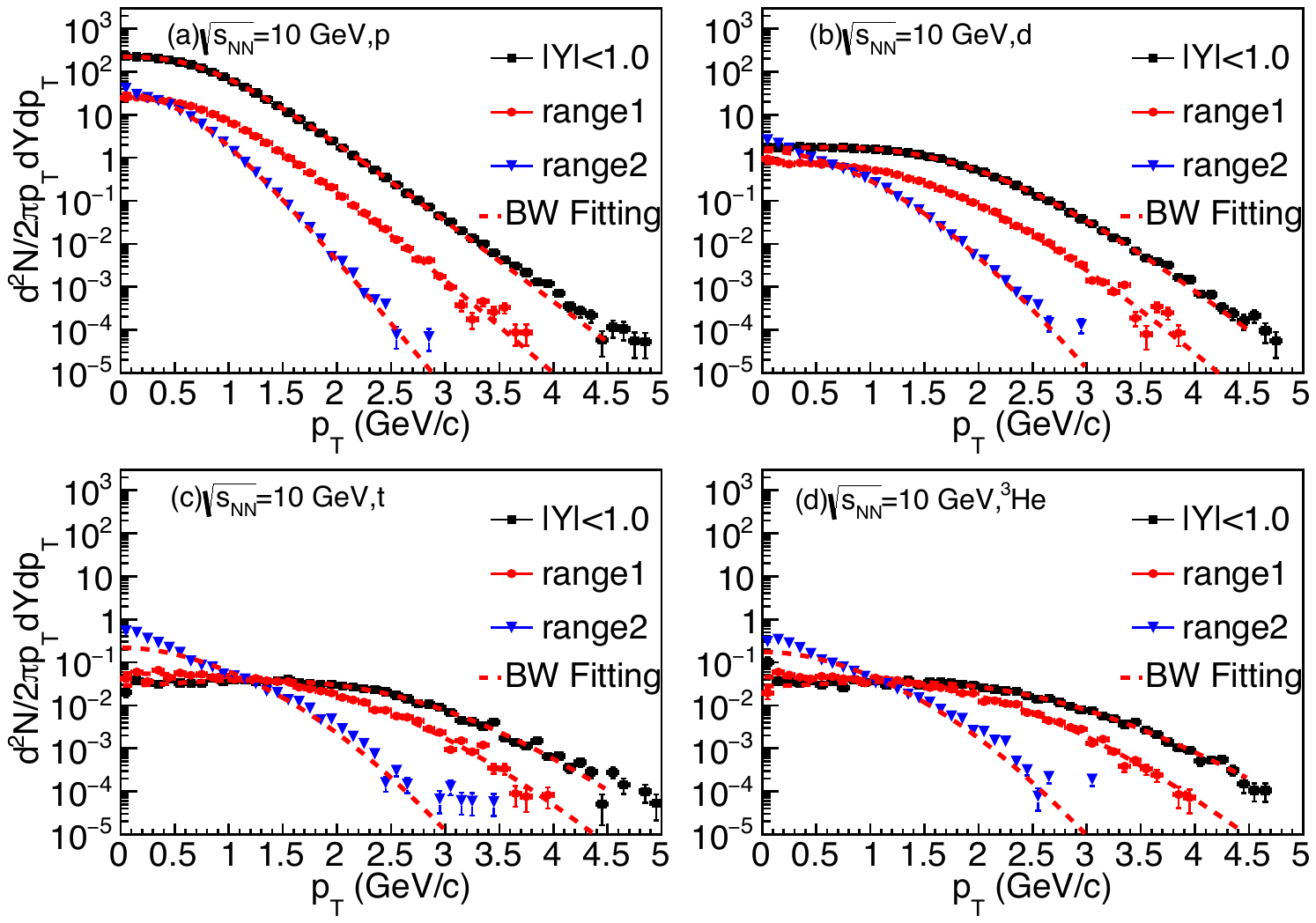}
\caption{(Color online) Transverse momentum P$_{T}$ distributions for proton (a), deuteron (b), triton (c), and $^{3}$He (d) within various rapidity cuts in central (b $<$ 3 fm) $^{197}$Au + $^{197}$Au collisions at c.m. energy $\sqrt{s_{\rm NN}}$ = 10 GeV with the UrQMD model. Red-dash curves are fitting lines with the Blast-wave (BW) model.}
\label{fig:fig3}
\end{figure}

\begin{figure}
\setlength{\abovecaptionskip}{0pt}
\setlength{\belowcaptionskip}{8pt}
\includegraphics[scale=0.59]{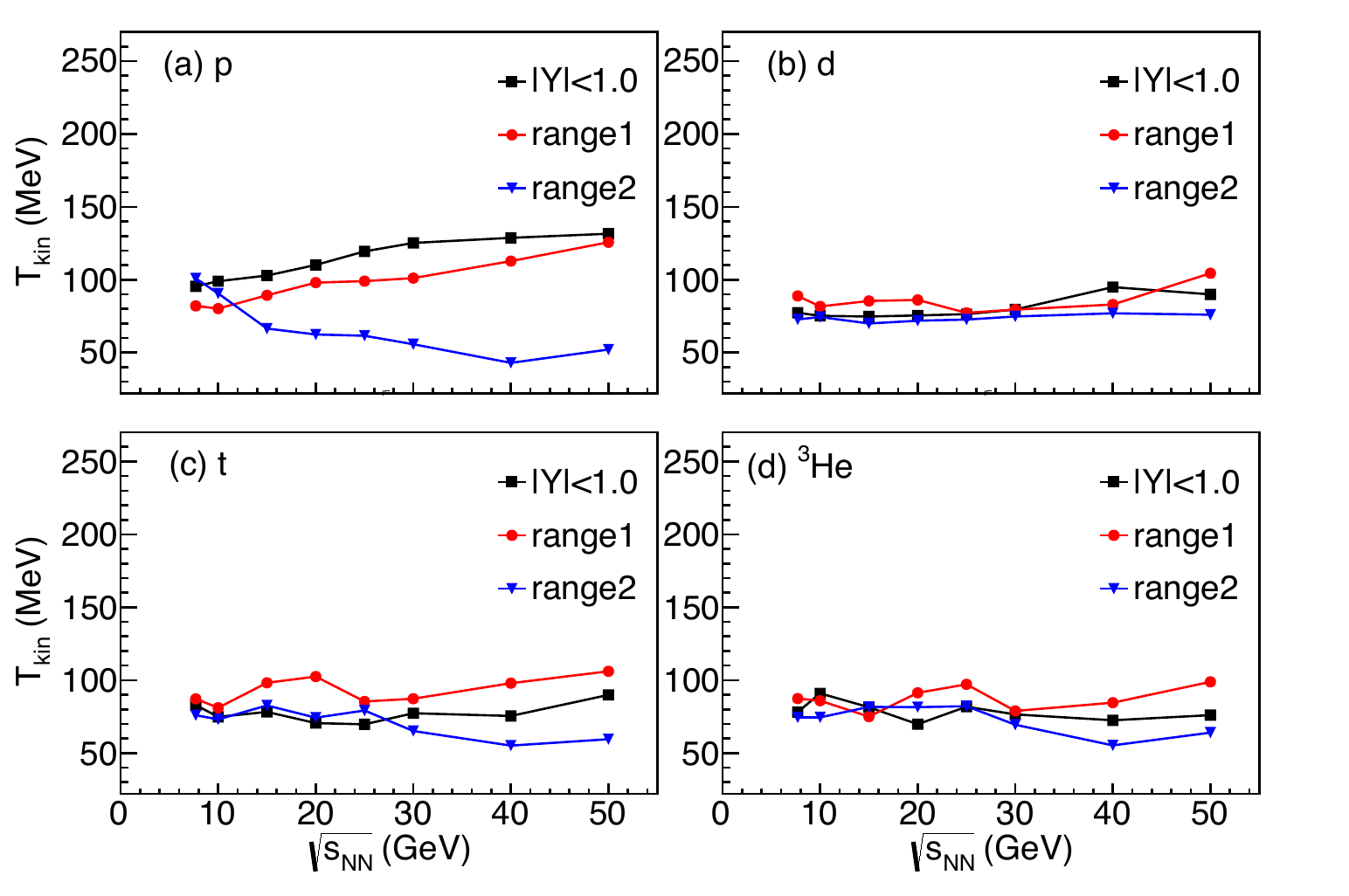}
\caption{(Color online) Kinetic temperature T$_{kin}$ for proton (a), deuteron (b), triton (c), and $^{3}$He (d) as a function of c.m. energy $\sqrt{s_{\rm NN}}$ within various rapidity cuts in central (b $<$ 3 fm) $^{197}$Au + $^{197}$Au collisions with the UrQMD model.}
\label{fig:fig4}
\end{figure}

\section{Results and discussion}
\label{TF2}
\subsection{The yield of light nuclei and kinetic temperature}

Firstly, we consider the rapidity density ($dN/dY$) as a function of $\sqrt{s_{\rm NN}}$ for two kinds of midrapidity cuts, namely $|{\rm Y}|<0.1$ and $|{\rm Y}|<0.3$, as displayed in Fig. ~\ref{fig:fig1}. The comparison is with the experimental results from the STAR collaboration \cite{BI09,LA17,JH18,JA19,LH20} as shown with circles, squares and inverted triangles. For both of $|{\rm Y}|<0.1$ and $|{\rm Y}|<0.3$,  $dN/dY$ of proton decreases as the increasing of  $\sqrt{s_{\rm NN}}$ and they are coincident with data. For the deuteron cases, they also decrease as the increasing of  $\sqrt{s_{\rm NN}}$. The values of $dN/dY$ of deuteron are less than those of data in $|{\rm Y}|<0.1$ window but it is coincident in $|{\rm Y}|<0.3$ window. However, for triton production, there is in a good agreement with that data within $|{\rm Y}|<0.1$ cut but  less than the data within $|{\rm Y}|<0.3$ cut. It indicates that even though in midrapidity region the yields of deuteron and triton are sensitively dependent of rapidity window. Also for triton yield, one can see that large fluctuations appear around 40 GeV within $|{\rm Y}|<0.1$ cut. The yield of $^{3}$He is  similar to the yield of triton. For antiparticle, the yield of anti-proton increases as $\sqrt{s_{\rm NN}}$ and is close to the yield of proton at high energy  as shown with the purple dot-dash lines. The reason here is that the baryon stopping dominates at low energies, while the pair production dominates at high energies, as stated in Ref.~\cite{LXF17,LH20}.

The rapidity distributions of proton, neutron, deuteron, triton, $^{3}$He, and $^{4}$He at various energies are plotted in Fig.~\ref{fig:fig2}. For distributions of proton and neutron, we can see that the shapes are narrow at low energies and the peak strengths are  higher, and  at higher energies, shapes around midrapidity are rather flat and values are less than those at lower energies. However, four-peak structure is obviously emerged for the other four light nuclei, namely deuteron, triton, $^3$He and $^4$He. 
The central two peaks of each above nucleus represent the target-like and projectile-like regions and the other two outer peaks display the spectator regions. In Fig.~\ref{fig:fig2} (c), values  at midrapidity  are not constant and the valleys are forming,  this leads to an obvious  rapidity dependent  deuteron's yield as discussed above. 
Furthermore, one can see that the yields of triton, $^{3}$He and $^{4}$He are very few around midrapidity and almost equal to zero. However, around target-like and projectile-like regions, the numbers are considerable. Thus the ratios at different regions of rapidity are worth to investigate. Therefore, for some calculations below, we extract Y$_{peak} $ and Y$_{1}$ as demarcation points from Fig.~\ref{fig:fig2}(c). For details, Y$_{peak}$ and Y$_{1}$  are defined as the first peak location and minimum location of rapidity at each $\sqrt{s_{NN}}$ on right hand side in Fig.~\ref{fig:fig2}(c), respectively. As an example, $Y_{peak}$ and $Y_1$ are marked in the insert for the 10 GeV case.  In this work, we set  $|{\rm Y}-{\rm Y}_{peak}|<0.05Y_{peak}$ and $|{\rm Y}+{\rm Y}_{peak}|<$0.05Y$_{peak}$ as `{\bf range1}', and  $|{\rm Y}_{1}|<|{\rm Y}|<|{\rm Y}_{1}+0.2|$ 
is treated as `{\bf range2}', i.e.  `{\bf range1}' corresponds to initial Au-like rapidity region and  `{\bf range2}' to spectator (cold nuclei) region.

Transverse momentum distributions for  proton,  deuteron, triton and $^{3}$He in different rapidity ranges at $\sqrt{s_{\rm NN}}$ = 10 GeV are shown in Fig.~\ref{fig:fig3}. The red-dash curves are fitted lines with the Blast-wave model \cite{ES93,FR04,DF20,Liu,LvM}, i.e., 
\begin{equation}
\frac{dN}{p_{T}dp_{T}} \propto \int_{0}^{R}r dr  m_{T}  I_{0}[\frac{p_{T}sinh(\rho)}{T_{kin}}] 
                                K_{1}[\frac{m_{T}cosh(\rho)}{T_{kin}}],                     
\label{BlastW}
\end{equation}
where 
$m_{\rm T}$ = $\sqrt{m^{2}+p_{\rm T}^{2}}$ is the transverse mass, $T_{kin}$ is kinetic temperature of particles at freeze-out, $r$ and $R$ are the radial position and the maximum radial position, respectively. $I_{0}$ and $K_{1}$ are the modified Bessel functions,  $\rho$ is the boost angle which is $tanh^{-1}[\beta(r)]$, $\beta(r) = \beta_{s}(r/R)^{\sigma}$ is a self-similar flow profile, $\beta_{s}$ is the surface velocity, and $\sigma$ is an index factor which is corresponding to  the shape of source. Kinetic temperatures are extracted by the Blast-wave fitting as shown in Fig.~\ref{fig:fig4}. One can see that the behaviors of temperature as a function of $\sqrt{s_{\rm NN}}$ for proton, deuteron, triton, and $^{3}{\rm He}$ in different rapidity regions. For protons in both rapidity regions of $|{\rm Y} |<1.0$ and `{\bf range1}', kinetic temperatures increase as $\sqrt{s_{\rm NN}}$. Since both of them are related to the fireball due to the flat or single-peak rapidity distribution, the higher the collision energy,  the hotter the proton's temperature. For the rapidity region of $|{\rm Y} |<1.0$ which is with respect to the central area in the fireball there is higher temperature than the one of `{\bf range1}' which is outside area of the fireball. This is as discussed in Section ~\ref{introduction}. 
In the `{\bf range2}' which are located in spectator region, the behavior of kinetic temperature is opposite. It indicates as increasing of collision energy, spectators pass through  so fast that they are got less excited. Thus kinetic temperature goes down  as increasing of energy. For the cases of deuteron, triton and $^{3}{\rm He}$, all their kinetic temperatures are about 80 MeV which are less than the one of proton for $|{\rm Y} |<1.0$ and `{\bf range1}'. It implies that the light nuclei such as  deuteron, triton and $^{3}{\rm He}$ are mostly coming from  non midrapidity region. From the rapidity distributions of  deuteron, triton and $^{3}{\rm He}$ in Fig.~\ref{fig:fig2}, it is expected that temperature  in the rapidity region of $|{\rm Y} |<1.0$ could be less than ones for `{\bf range1}' in
Fig.~\ref{fig:fig2}(c), because  the midrapidity region of $|{\rm Y} |<1.0$ for these nuclei is not really midrapidity particles, but just the tailed particles of Au-like region. 

\subsection{The ratios of light nuclei}

\begin{figure}
\setlength{\abovecaptionskip}{0pt}
\setlength{\belowcaptionskip}{8pt}
\includegraphics[scale=0.58]{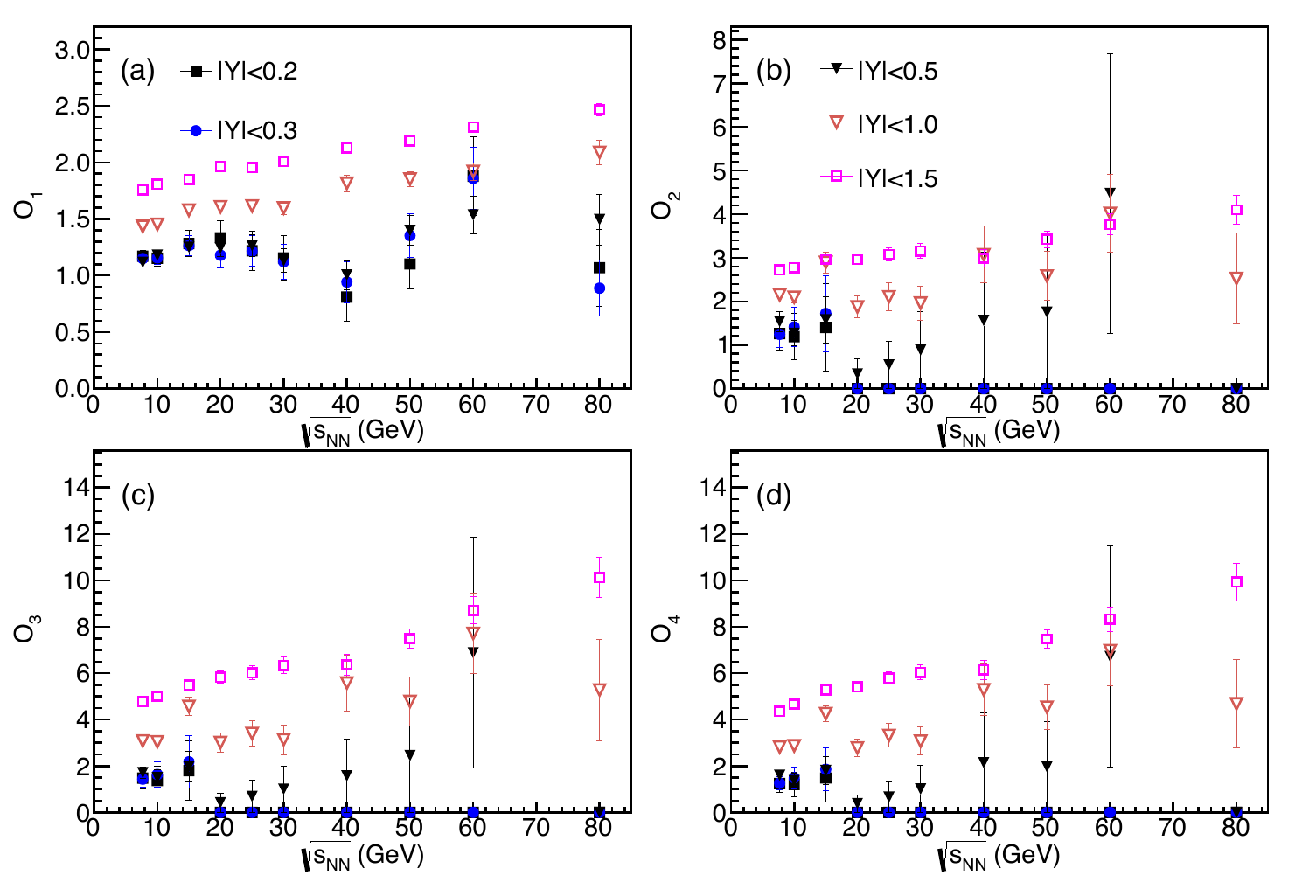}
\caption{(Color online) Ratios O${_1}$, O${_2}$, O${_3}$, and O${_4}$ as functions of $\sqrt{s_{\rm NN}}$ for various midrapidity cuts at central (b $<$ 3 fm) $^{197}$Au + $^{197}$Au collisions within the UrQMD model.}
\label{fig:fig5}
\end{figure}

\begin{figure}
\setlength{\abovecaptionskip}{0pt}
\setlength{\belowcaptionskip}{8pt}
\includegraphics[scale=1.2]{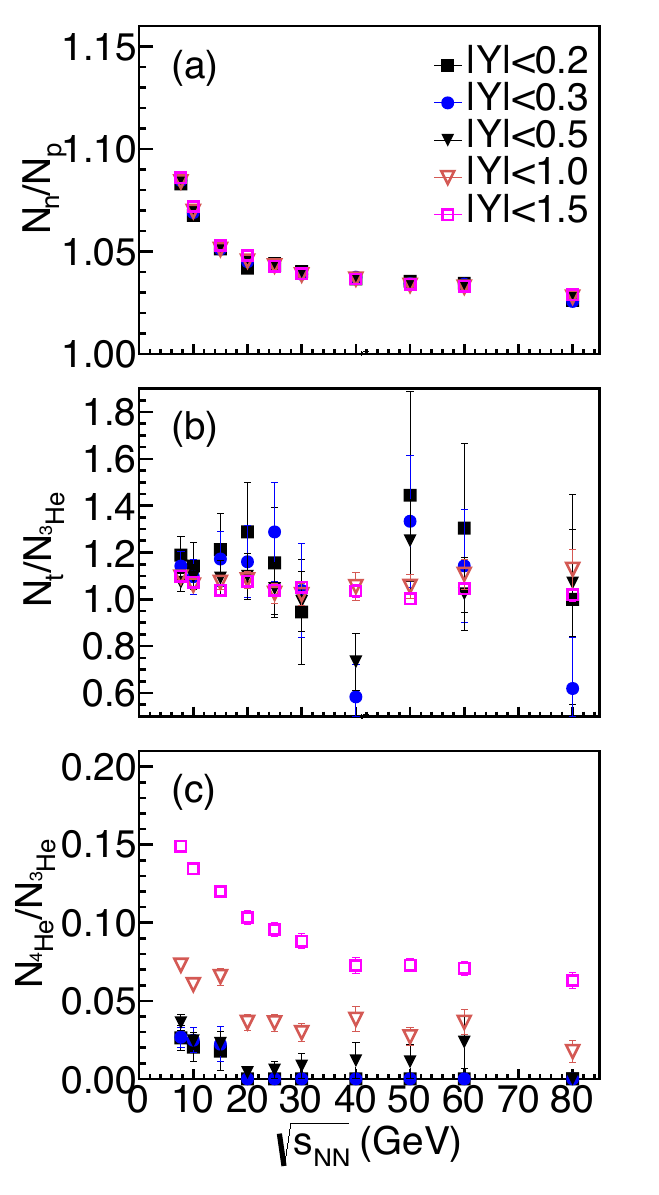}
\caption{(Color online) Ratios of $N_n/N_p$, $N_t/N_{^{3}{\rm He}}$
 and $N_{^{4}{\rm He}}/N_{^{3}{\rm He}}$ as functions of $\sqrt{s_{\rm NN}}$ with various midrapidity cuts at central (b $<$ 3 fm) $^{197}$Au + $^{197}$Au collisions within the UrQMD model.}
\label{fig:fig6}
\end{figure}

Ratios of $O_{1}$, $O_{2}$, $O_{3}$, and $O_{4}$ as functions of $\sqrt{s_{\rm NN}}$ with different midrapidity cuts are shown in Fig.~\ref{fig:fig5}. For $O_{1}$ which presents the neutron density fluctuation shown in Fig.~\ref{fig:fig5}(a), it seems to show a slight enhancement arising
around 20 GeV and another broad peak emerges at 60 GeV with small midrapidity cuts as displayed with black solid-squares, blue solid-circles and black inverted triangles. However, the UrQMD model which we are using does not include    first-order or second-order phase transition mechanisms \cite{SA98,HP08,JW20}. In this context, the enhancement around 20 GeV may not indicate the CEP.  Also in the present UrQMD model, the cascade mode of the UrQMD model might be simply for the real simulations of heavy-ion collisions at very high energy (as above 40 GeV) because only the hadrons and strings are taken into consideration, even though we can reproduce appropriate light nuclei yields as depicted in Fig.~\ref{fig:fig1}. Thus the peak at 60 GeV should be also treated with caution. In other panels for  $O_{2}$, $O_{3}$, and $O_{4}$, ratios are around 1-2 below 20 GeV and tend to zero as higher energies   within midrapidity cuts because of the negligible   $^{4}$He production beyond 20 GeV. 
Regardless, in Fig.~\ref{fig:fig5}(a) to \ref{fig:fig5}(d)  for higher midrapidity cuts with $|$Y$|<$1.5, the ratios increase as  $\sqrt{s_{\rm NN}}$.  In particular, the ratios $O_3$ and  $O_4$ are close to each other in all rapidity regions due to the similar production yield of triton and $^3$He which can be seen from next figure (Fig.~\ref{fig:fig6}(b)).

For the ratio of $N_n/N_p$ which is usually taken as a sensitive probe to neutron skin \cite{SunXY,YanTZ,LiWJ}, we can see from Fig.~\ref{fig:fig6}(a) that for all midrapidity cuts  the ratios decrease as the increasing of energy and all the ratios are the same at a given energy since neutrons and protons are basically coming from a single midrapidity source (participants). The ratios of triton to $^{3}$He in Fig.~\ref{fig:fig6}(b) for midrapidity cuts with $|$Y$|$$<$1.0 and $|$Y$|$$<$1.5 are showing nearly constant value as $\sqrt{s_{\rm NN}}$ increases. According the nucleon component, one may expect $N_{t}/N_{^{3}{\rm He}}$ has the same value as $N_n/N_p$. However, only up till high energy region, ratio of triton to $^{3}$He has the similar ratio of $N_n/N_p$. In low energy region, ratio of $N_n/N_p$ is affected by the initial isospin asymmetry which is 118/79 for $^{197}$Au. And also such initial isospin asymmetry seems to affect  $N_{^{4}{\rm He}}/N_{^{3}{\rm He}}$  in low energy region. With the increasing of $\sqrt{s_{NN}}$, 
$N_{^{4}{\rm He}}/N_{^{3}{\rm He}}$ tends to the constant for each given rapidity window, indicating the relative production rate for $N_{^{4}{\rm He}}/N_{^{3}{\rm He}}$ keeping close at higher energies. In addition, the larger the rapidity region, the higher the $^{4}$He/$^{3}$He, which reflects that heavier light nuclei could be preferentially formed at larger rapidity.

\begin{figure}
\setlength{\abovecaptionskip}{0pt}
\setlength{\belowcaptionskip}{8pt}
\includegraphics[scale=0.58]{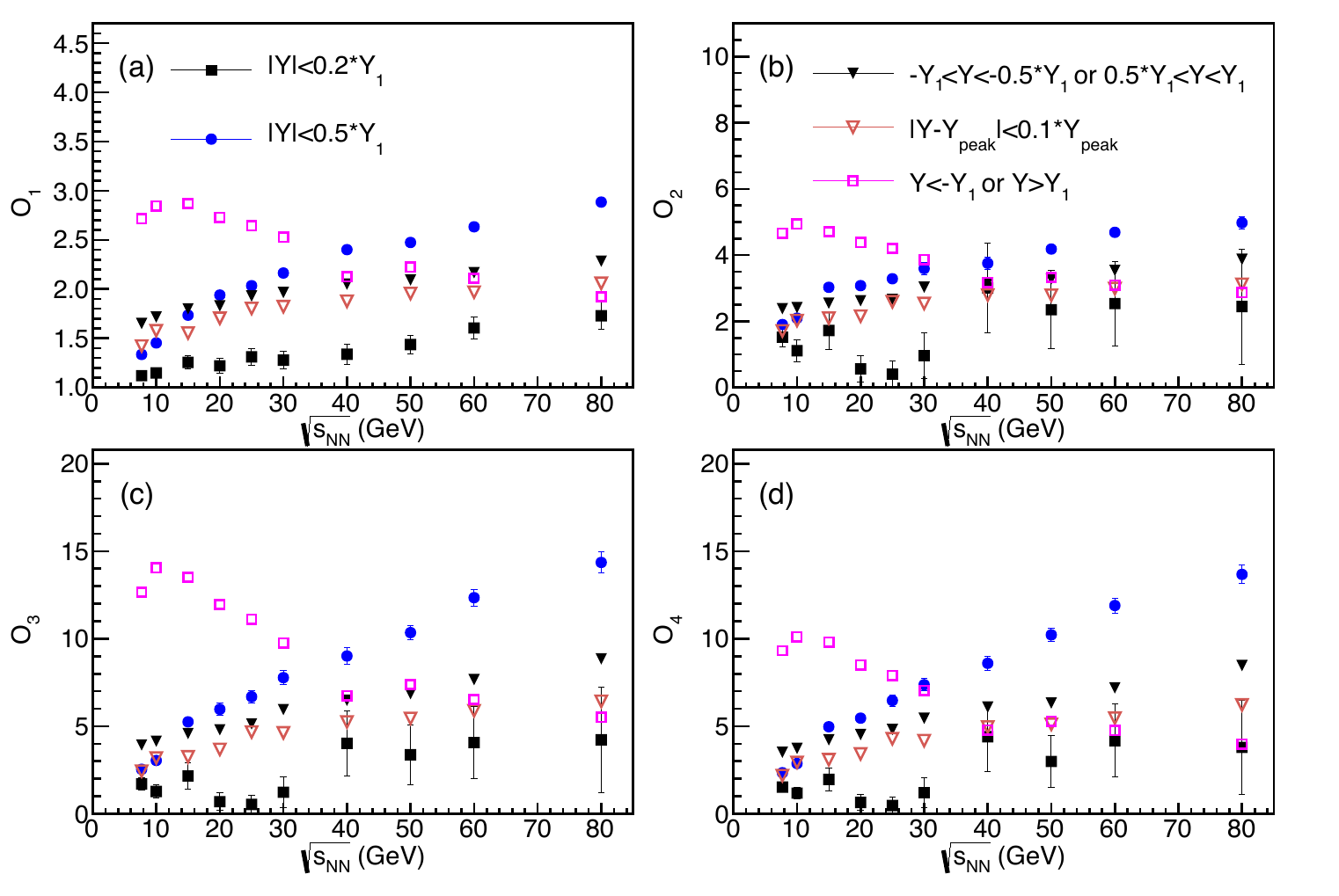}
\caption{(Color online) Ratios O${_1}$, O${_2}$, O${_3}$, and O${_4}$ as functions of $\sqrt{s_{\rm NN}}$ for various midrapidity regions at  central (b $<$ 3 fm) Au + Au collisions within the UrQMD model.}
\label{fig:fig7}
\end{figure}

\begin{figure}
\setlength{\abovecaptionskip}{0pt}
\setlength{\belowcaptionskip}{8pt}
\includegraphics[scale=1.2]{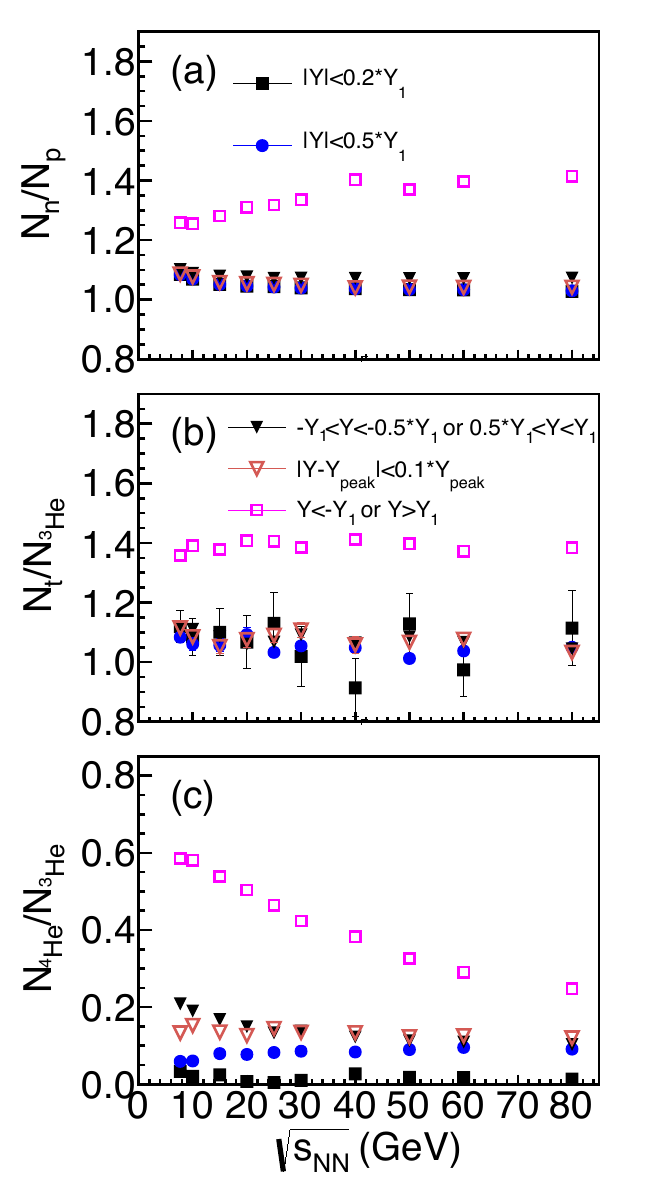}
\caption{(Color online) Ratios of $N_n/N_p$,  $N_t/N_{^{3}{\rm He}}$ and $N_{^{4}{\rm He}}/N_{^{3}{\rm He}}$ as functions of $\sqrt{s_{\rm NN}}$ for various midrapidity regions at central (b $<$ 3 fm) $^{197}$Au + $^{197}$Au collisions within the UrQMD model.}
\label{fig:fig8}
\end{figure}

As mentioned above, we separated the rapidity into various regions with boundary values of Y$_{peak} $ and Y$_{1}$. In Fig.~\ref{fig:fig7}(a), by comparing  $O_{1}$ ratios among Inner region of $|Y|<$0.2Y$_{1}$ ($O_I$), Middle region of $|$Y$|<$0.5Y$_{1}$ ($O_M$), and Outside region of 0.5Y$_{1}$ $<|$Y$|< Y_{1}$ ($O_O$) inside the entire region of -Y$_{1}<$Y$<$Y$_{1}$,  we found $O_I<O_O<O_M$. 
It indicates that in Inner region, the matter is more uniform and would be less of neutron density fluctuations.  And for Middle region, there are more  kinds of  particle production than those in Outside region and then less  uniform than the Inner one, thus it has more neutron density fluctuations with higher ratios of $O_{1}$.
The ratios of Outside one are close to the ratios in target-like and projectile-like region which is $|$Y-Y$_{peak}|<$0.1Y$_{peak}$. The purple empty-squares present that the ratios in the spectator regions  ($|Y|>Y_{1}$) 
increase a little bit up till 10 GeV and then decrease as energy increases. The reason could be  that as the increasing of energy up till 10 GeV, the spectators  break up and then the neutron density fluctuation increases. Then as energy goes higher, there are more free nucleons emitted,  and then the neutron density fluctuation becomes less. One more thing we should notice is that ratios of $O_{2}$, $O_{3}$ and $O_{4}$ 
have very similar trends to each other, which can be attributed to the dominant  $^{4}$He in these ratios. In addition,  $O_3$ can be obtained by $O_1 \times O_2$ or either by $O_4 \times  N_t/N_{^{3}{\rm He}}$.

Furthermore, ratios of $N_n/N_p$, $N_t/N_{^{3}{\rm He}}$ and $N_{^{4}{\rm He}}/N_{^{3}{\rm He}}$ in various rapidity regions are shown in Fig.~\ref{fig:fig8}. In Fig. \ref{fig:fig8}(a), except for the one of purple symbols in the spectator region, the ratios of $N_n/N_p$ in all other regions are the same. For the purple symbols, as energy increases they tend to 1.4 which are close to the initial isospin asymmetry 118/79 $\approx$ 1.49. The ratio of $N_t/N_{^{3}{\rm He}}$ in the spectator region keeps at value of 1.4 in Fig. ~\ref{fig:fig8}(b). Both ratios of $N_n/N_p$ and $N_t/N_{^{3}{\rm He}}$ are the same as discussed above, which indicates that $N_t/N_{^{3}{\rm He}}$ could be taken as a reasonable approximation to $N_n/N_p$ as medium interaction is not very strong and then suggested  as a neutron-skin probe at Fermi energy \cite{DaiZT}. For   $N_{^{4}{\rm He}}/N_{^{3}{\rm He}}$ in Fig.~\ref{fig:fig8}(c),
the values in the spectator region drop quickly with $\sqrt{s_{NN}}$ but others show  rather flat dependence of $\sqrt{s_{NN}}$.

\section{Conclusion}
\label{summary}

In this work, we extracted different single ratios and combined ratios of light nuclei by naive coalescence approach in the framework of UrQMD model. By comparing with  the data, the yields of light nuclei seem reasonable. Meanwhile, kinetic temperatures of proton, deuteron, triton and $^3$He  in different rapidity regions are extracted based on the Blast-wave model assumption. For the combined ratio ${N_{p}N_{t}}/{N_{d}^{2}}$ in midrapidity region which is thought to be sensitive to neutron density fluctuation, it seems that a slight enhancement is observed  around 20 GeV, however, it should not be over-explained as the sign of critical end point due to the physics ingredient of the UrQMD, which could be arisen  by other mechanism or due to the less precision of this naive coalescence approach. Other combined ratios involved $^4$He are also checked, and found very similar behavior due to the dominant role of $^4$He. Except for the midrapidity particles and their ratios, we also consider ratios of light nuclei in other rapidity regions. 
Based on the present result, it indicates that there are lots of information we can learn from the outside midrapidity while a suitable model is further expected.

\vspace{.5cm}
{\bf Acknowledgments.---}  
This work was partially supported by
the National Natural Science Foundation of China under Contract
Nos. 11890714, 11421505,  11947217 and 2018YFA0404404, China Postdoctoral Science Foundation Grant  No. 2019M661332, Postdoctoral Innovative Talent Program of China No. BX20200098, the Strategic Priority Research Program of the CAS under
Grant No. XDB34030200  and XDB16, and the Key Research Program of Frontier Science of CAS under Grant NO. QYZDJ-SSW-SLH002.

\end{CJK*}
\end{document}